\documentclass[10pt,oneside]{article}
\usepackage[T1]{fontenc}
\usepackage[utf8]{inputenc}
\usepackage[english]{babel}
\usepackage{amsmath,amssymb,amsfonts}
\usepackage{graphicx}
\usepackage{float}
\usepackage{flafter}
\usepackage[section]{placeins}
\usepackage{xcolor}
\usepackage{geometry}
\usepackage[numbers,sort&compress]{natbib}
\usepackage{authblk}
\usepackage{appendix}
\usepackage{hyperref}
\usepackage{wasysym}

\begin{document}
\title{Magnetoelastic control of quantum correlations and field sensitivity
in a spin-1/2 Heisenberg dimer}

\author[1]{E. W. B. de Souza}
\author[1]{Moises Rojas}
\author[1]{Onofre Rojas}

\affil[1]{Department of Physics, Institute of Natural Sciences,
Federal University of Lavras, Lavras, Minas Gerais, Brazil}
\date{}

\maketitle

\begin{abstract}
We investigate the thermodynamic and quantum properties of a magnetoelastic
spin-1/2 Heisenberg dimer, where the exchange interaction depends
on the dimer displacement. By combining an exact treatment of the
spin sector with a harmonic description of the vibrational degree
of freedom, we obtain an effective model in which each spin configuration
is associated with a distinct vibrational mode, leading to a non-factorizable
partition function. We analyze the thermal behavior and identify regimes
corresponding to entangled and fully polarized states. Quantum correlations
are analyzed through concurrence and local quantum uncertainty, showing
that while entanglement is rapidly suppressed by temperature, nonclassical
correlations persist over a broader range due to the competition between
spin sectors. We further examine the magnetic Fisher information,
which provides a measure of the sensitivity of the system to the external
magnetic field. Its behavior reveals enhanced response in crossover
regions where magnetoelastic effects induce strong redistribution
of the level populations. Our results demonstrate that magnetoelastic
coupling plays a central role in controlling both quantum correlations
and magnetic response, establishing a direct link between entanglement,
nonclassical correlations, and thermodynamic sensitivity in coupled
spin-dimer systems.
\end{abstract}


\section{Introduction}
\label{sec:introduction}

Low-dimensional quantum spin systems have attracted considerable attention
due to the emergence of nonclassical correlations such as thermal
entanglement \citep{Amico,Horodecki,Guhne} and quantum coherence\citep{HuHu,streltsov,ref2-C_l1,kraft,ref1-C_l1,filgueiras}.
In particular, Heisenberg dimers constitute paradigmatic systems for
investigating quantum correlations in thermal equilibrium, since their
spectra can be controlled through external magnetic fields, exchange
interactions, and anisotropies \citep{wootters,hill}. Owing to their
analytical tractability and direct connection with magnetic compounds,
spin dimers have been extensively employed in the study of thermal
entanglement and related quantum properties \citep{Arnesen,xWang}.
Moreover, the competition between different magnetic sectors may induce
nontrivial thermal redistribution of populations, strongly affecting
entanglement and other quantum observables.

Although concurrence provides a well-established measure of bipartite
entanglement \citep{wootters,hill}, it does not capture all forms
of quantum correlations present in mixed thermal states. In this context,
quantities beyond entanglement have attracted increasing attention
in low-dimensional spin systems. Among them, the local quantum uncertainty
(LQU), introduced by Girolami et al. \citep{Girolami}, provides a
useful measure of nonclassical correlations based on the Wigner-Yanase
skew information. An important feature of the LQU is that it may remain
finite even in thermal regimes where the concurrence vanishes, revealing
quantum correlations beyond entanglement. Moreover, the LQU is closely
related to quantum metrology and parameter-estimation protocols \citep{Girolami,Girolami14,Metrolgy}.
Recent studies have also explored the interplay between LQU, local
quantum Fisher information, and discord-type correlations in Heisenberg
spin systems with anisotropic interactions and thermal fluctuations
\citep{pfwei,Yurischev,Yurischev25}, emphasizing their usefulness
for characterizing quantum correlations in low-dimensional magnetic
systems.

Another quantity of growing interest is the Fisher information, which
characterizes the sensitivity of a system to external perturbations
and plays a central role in quantum estimation theory and quantum
metrology\citep{Metrolgy,Braunstein,paris}. In magnetic spin systems,
the Fisher information is directly connected with field-induced redistribution
of thermal populations and magnetic fluctuations, providing an information-theoretic
measure of the magnetic response. In low-dimensional systems, such
enhanced sensitivity is often amplified by thermal and quantum fluctuations,
making the Fisher information a useful tool for investigating the
interplay between magnetic response and quantum correlations. Recent
studies have further explored the role of quantum Fisher information
and related metrological quantifiers in Heisenberg spin models, revealing
their usefulness for detecting thermal correlations, critical behavior,
and enhanced parameter sensitivity in low-dimensional quantum magnets
\citep{pfwei,Yurischev,Yurischev25,Fedorova,Leny,Anouz,Phsy-E,zad-rojas}.

Atomic vibrations in crystalline materials may strongly influence
magnetic ordering and, conversely, magnetic interactions may induce
lattice deformations through magnetoelastic coupling \citep{Henriques,massimino,boubcheur}.
This interplay gives rise to magnetostriction effects and modifies
the thermodynamic and magnetic properties of the system. Some compounds,
such as $\mathrm{Tb_{0.3}Dy_{0.7}Fe_{1.9}}$, may exhibit giant magnetostriction
\citep{Armstrong}, while magnetoelastic effects also play an important
role in ferromagnetic shape-memory alloys such as $\mathrm{Ni_{2}MnGa}$
\citep{tickle,Tickle-jmmm}. More recently, renewed interest in magnetoelastic
phenomena has emerged in quantum materials and low-dimensional magnetic
systems, including molecular spin qubits \citep{Dunstan}, itinerant
ferromagnets \citep{Marques}, magnon-phonon hybrid systems \citep{Choe},
and strain-controlled two-dimensional magnetic materials \citep{Teh}.
Owing to the coupling between spin and lattice degrees of freedom,
the theoretical description of magnetoelastic systems remains challenging,
particularly in the presence of quantum and thermal fluctuations.

Magnetoelastic effects have been previously investigated in several
exactly solvable spin models, including mixed spin-$(1/2,S)$ Ising
systems on decorated lattices \citep{strecka12,strecka19}. In such
models, local canonical transformations allow the magnetic and lattice
degrees of freedom to be partially decoupled, leading to effective
interactions induced by lattice distortions \citep{enting}. Related
effects have also been explored in XXZ-Ising diamond chains \citep{Nayara}.
In contrast to those classical or semi-classical systems, the present
work focuses on a fully quantum magnetoelastic Heisenberg dimer, where
lattice-induced renormalization directly affects thermal populations
and consequently modifies entanglement, nonclassical correlations,
and magnetic response.

Motivated by these considerations, we investigate a magnetoelastic
spin-1/2 XXZ Heisenberg dimer in which the exchange interactions depend
explicitly on the lattice displacement. Combining an exact treatment
of the spin sector with an effective harmonic description of the vibrational
degree of freedom, we analyze how magnetoelastic renormalization affects
thermal populations, quantum correlations, and magnetic response.
In particular, we investigate the interplay between concurrence, local
quantum uncertainty, and magnetic Fisher information, emphasizing
the role of vibrationally induced population redistribution in the
thermal behavior of the system.

The paper is organized as follows. In Sec.~\ref{sec:model} we introduce the magnetoelastic
Heisenberg dimer and derive its effective spectrum, including the
ground-state phase diagram. In Sec.~\ref{sec:thermal} we discuss the thermodynamic
properties, thermal populations, and magnetoelastic fluctuations.
In Sec.~\ref{sec:quantum-response} we analyze quantum correlations and magnetic response through
concurrence, local quantum uncertainty, and magnetic Fisher information.
Finally, Sec.~\ref{sec:conclusions} summarizes our conclusions.

\section{Heisenberg dimer with magnetoelastic coupling}
\label{sec:model}

We consider a spin-1/2 Heisenberg dimer composed of two localized
spins, denoted by operators $\mathbf{S}_{1}$ and $\mathbf{S}_{2}$,
subjected to an external magnetic field $B$. The spins are hosted
by two atoms whose equilibrium separation is $r_{0}$. In addition
to their magnetic degrees of freedom, the atoms are allowed to perform
small oscillations around their equilibrium positions, giving rise
to dimer (phononic) fluctuations that couple to the spin sector.

The spin-1/2 Heisenberg dimer Hamiltonian provides a minimal description
of several dimerized magnetic compounds and quantum spin-gap systems,
including the Shastry-Sutherland material $\mathrm{SrCu_{2}(BO_{3})_{2}}$
\citep{Kageyama,Miyahara}. In these systems, the dominant intradimer
exchange interaction justifies the use of an isolated dimer approximation,
while anisotropic extensions of the model are relevant for describing
field-induced excitations and anisotropic magnetic responses \citep{Nojiri}.

\subsection{Spin-phonon Hamiltonian and relative coordinate}
\label{sec:spin-phonon}

\begin{figure}[!h]
\centering
\includegraphics[width=0.5\linewidth]{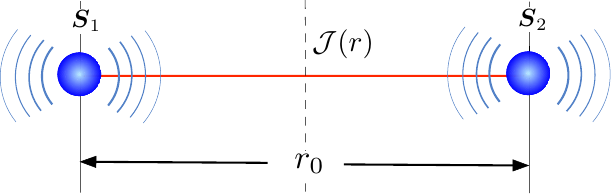}\caption{\label{dimer} Schematic representation of a dimer spin oscillating
around its equilibrium position $r_{0}$.}
\end{figure}

The two spins are attached to atoms forming a dimer, whose equilibrium
separation is $r_{0}$ (see Fig.~\ref{dimer}). The atomic displacements
from equilibrium are denoted by $\mathbf{r}_{1}$ and $\mathbf{r}_{2}$.
We restrict the motion to the dimer axis and introduce a scalar relative
displacement $r$, defined as the deviation from the equilibrium separation.
Thus, the instantaneous distance between the spins is $r_{0}+r$,
and $r=0$ corresponds to equilibrium.

The conjugate momenta are $\mathbf{p}_{1}$ and $\mathbf{p}_{2}$.
After separating the centre-of-mass motion, which is assumed to be
decoupled, the dynamics is described solely in terms of the relative
coordinate $r$ and its conjugate momentum $p$.

We consider a spin-1/2 Heisenberg dimer coupled to a single vibrational
mode associated with this internal relative motion. The total Hamiltonian
reads
\begin{equation}
H=H^{(\mathrm{me})}+H^{(\mathrm{ph})},\label{eq:H-tot}
\end{equation}
where the vibrational (phonon-like) contribution is
\begin{equation}
H^{(\mathrm{ph})}=\frac{\boldsymbol{p}^{2}}{2m}+\frac{\kappa}{2}r^{2},\label{eq:H-ph}
\end{equation}
Here $m$ is the effective mass associated with the relative motion,
$p$ is the conjugate momentum of $r$, and $\kappa$ is the elastic
constant describing the restoring force around the equilibrium separation.
This harmonic form corresponds to small oscillations of the two atoms
forming the dimer.

The magnetic Hamiltonian depends explicitly on the relative displacement:
\begin{alignat}{1}
H^{(\mathrm{me})}(r)=  -\mathcal{J}_{x}(r)\bigl(S^{x}_{1}S^{x}_{2}+S^{y}_{1}S^{y}_{2}\bigr)-\mathcal{J}_{z}(r)\,S^{z}_{1}S^{z}_{2}
 -g\mu_{B}B\bigl(S^{z}_{1}+S^{z}_{2}\bigr).\label{eq:H-me}
\end{alignat}

To incorporate magnetoelastic effects within the dimer, we assume
that the exchange couplings depend on the relative displacement. For
small oscillations around equilibrium, $r_{0}$, the Taylor expansion
around relative displacement up to second order becomes
\begin{equation}
\mathcal{J}_{\alpha}(r)=\mathcal{J}_{\alpha}(0)+\tfrac{\partial\mathcal{J}_{\alpha}(r)}{\partial r}|_{r=0}r+\tfrac{\partial^{2}\mathcal{J}_{\alpha}(r)}{2\partial r^{2}}|_{r=0}r^{2}+\mathcal{O}(r^{3}),
\end{equation}
with $\alpha=\{x,z\}$, and defining $\mathcal{J}_{\alpha}(0)=J_{\alpha}$,
$\left.\tfrac{\partial\mathcal{J}_{\alpha}(r)}{\partial r}\right|_{r=0}
=-\gamma_{\alpha}J_{\alpha}$, and, for simplicity, assuming
$\left.\tfrac{\partial^{2}\mathcal{J}_{\alpha}(r)}{\partial r^{2}}\right|_{r=0}
=\gamma^{2}_{\alpha}J_{\alpha}$.
\begin{equation}
\mathcal{J}_{\alpha}(r)\approx J_{\alpha}\left[1-\gamma_{\alpha}r+\tfrac{1}{2}\gamma^{2}_{\alpha}r^{2}\right].\label{eq:2nd-corr}
\end{equation}
The condition $|\gamma_{\alpha}|\Delta r\ll1$, where
$\Delta r$ characterizes the fluctuation width of the relative
displacement, is a useful but only partial check; the sector-resolved
small-displacement criterion is given below. The linear
term describes the leading magnetoelastic coupling, while the quadratic
correction becomes relevant when the amplitude of the relative motion
is not negligible.

The expansion must be tested in every thermally relevant spin sector
using
\begin{equation}
\eta(T)=\max_{j,\alpha}\left\{|\gamma_{\alpha}|
\sqrt{\left\langle r^2\right\rangle_{j,T}}\right\},
\label{eq:eta-validity}
\end{equation}
where $\langle r^2\rangle_{j,T}$ is the conditional mean-square
displacement; its explicit form is derived in
Sec.~\ref{sec:spin-spectrum} and Sec.~\ref{sec:reduction}. The requirement
$\eta(T)\ll1$ becomes more restrictive as $|\gamma_\alpha|$ or
temperature increases, or a sector softens. Our results are restricted
to this window. Beyond it, anharmonic terms must be retained, for
example through self-consistent harmonic methods with
fluctuation-renormalized parameters. Quantum formulations also include
higher-order quantum and thermal spin fluctuations
\citep{VillelaQSCHA} and are particularly useful in frustrated or
low-dimensional many-body systems.

This coupling provides a direct mechanism by which the internal vibrational
motion of the dimer modulates the exchange interaction. As shown below,
the quadratic structure allows each spin sector to be mapped onto
an effective harmonic oscillator, leading to a renormalized spectrum.
The resulting thermal occupation of the eigenstates, particularly
those associated with entangled states, establishes a direct link
between magnetoelastic coupling and quantum correlations.

\subsection{Spin eigenstates and spectrum at fixed displacement}
\label{sec:spin-spectrum}

For a fixed value of the relative coordinate $r$, the magnetoelastic
Hamiltonian defined in \eqref{eq:H-me} remains diagonal in the same
eigenstates as the rigid XXZ dimer. This follows from the fact that
the spin operators are unchanged, and the displacement r only modulates
the exchange couplings. 

We employ the standard two-spin basis composed of the triplet and
singlet states,
\begin{alignat}{1}
|\varphi_{1}\rangle=  |^{+}_{+}\rangle,\qquad
|\varphi_{2}\rangle=  \tfrac{1}{\sqrt{2}}\left(|^{+}_{-}\rangle+|{}^{-}_{+}\rangle\right),\qquad
|\varphi_{3}\rangle=  \tfrac{1}{\sqrt{2}}\left(|^{+}_{-}\rangle-|{}^{-}_{+}\rangle\right),\qquad
|\varphi_{4}\rangle=  |^{-}_{-}\rangle.\label{eq:phi-st}
\end{alignat}
In this basis, the Hamiltonian is diagonal, and the eigenvalues become
explicit functions of the relative displacement:
\begin{equation}
E_{1}(r)  =-\frac{\mathcal{J}_{z}(r)}{4}-g\mu_{B}B,\quad
E_{2}(r)  =\frac{\mathcal{J}_{z}(r)}{4}-\frac{\mathcal{J}_{x}(r)}{2},\quad
E_{3}(r)  =\frac{\mathcal{J}_{z}(r)}{4}+\frac{\mathcal{J}_{x}(r)}{2},\quad
E_{4}(r)  =-\frac{\mathcal{J}_{z}(r)}{4}+g\mu_{B}B.
\end{equation}
To make explicit the magnetoelastic corrections, we expand the energies
up to second order in $r$, using the expansion introduced in
Eq.~\eqref{eq:2nd-corr}. This yields
\begin{equation}
E_{j}(r)=\mathfrak{e}^{(0)}_{j}-r\,\mathfrak{e}^{(1)}_{j}+r^{2}\,\mathfrak{e}^{(2)}_{j}.\label{eq:ek}
\end{equation}
The zeroth-order contribution corresponds to the rigid dimer spectrum
evaluated at equilibrium,
\begin{alignat}{1}
\mathfrak{e}^{(0)}_{1}=  -\frac{J_{z}}{4}-\mu_{B}gB,\qquad
\mathfrak{e}^{(0)}_{2}=  \frac{J_{z}}{4}-\frac{J_{x}}{2},\qquad
\mathfrak{e}^{(0)}_{3}=  \frac{J_{z}}{4}+\frac{J_{x}}{2},\qquad
\mathfrak{e}^{(0)}_{4}=  -\frac{J_{z}}{4}+\mu_{B}gB.\label{eq:e04}
\end{alignat}
The linear coefficients, which encode the leading coupling between
the internal vibrational motion and the spin degrees of freedom, are
given by 
\begin{alignat}{1}
\mathfrak{e}^{(1)}_{1}=  -\frac{J_{z}}{4}\gamma_{z},\qquad
\mathfrak{e}^{(1)}_{2}=  \frac{J_{z}}{4}\gamma_{z}-\frac{J_{x}}{2}\gamma_{x},\qquad
\mathfrak{e}^{(1)}_{3}=  \frac{J_{z}}{4}\gamma_{z}+\frac{J_{x}}{2}\gamma_{x},\qquad
\mathfrak{e}^{(1)}_{4}=  -\frac{J_{z}}{4}\gamma_{z}.\label{eq:e14}
\end{alignat}
The quadratic coefficients $\mathfrak{e}^{(2)}_{j}$ read 
\begin{alignat}{1}
\mathfrak{e}^{(2)}_{1}=  -\frac{J_{z}}{8}\gamma^{2}_{z},\qquad
\mathfrak{e}^{(2)}_{2}=  \frac{J_{z}}{8}\gamma^{2}_{z}-\frac{J_{x}}{4}\gamma^{2}_{x},\qquad
\mathfrak{e}^{(2)}_{3}=  \frac{J_{z}}{8}\gamma^{2}_{z}+\frac{J_{x}}{4}\gamma^{2}_{x},\qquad
\mathfrak{e}^{(2)}_{4}=  -\frac{J_{z}}{8}\gamma^{2}_{z}.\label{eq:e24}
\end{alignat}

These expressions show that each spin sector experiences a distinct
effective potential as a function of the relative displacement. The
linear term describes the leading magnetoelastic coupling, while the
quadratic term becomes relevant when the amplitude of the internal
vibrational motion is not negligibly small.

Since the eigenstates remain identical to those of the rigid dimer,
the coupling to the vibrational degree of freedom enters exclusively
through the $r$-dependence of the eigenvalues. As shown in
Sec.~\ref{sec:reduction}, this structure allows one to eliminate the
vibrational coordinate by treating each spin sector as a displaced
harmonic oscillator. The resulting procedure leads to a renormalization
of the energy spectrum and, in general, to a modification of the effective
vibrational frequency.

\subsection{Magnetoelastic reduction and effective spectrum}
\label{sec:reduction}

Using the quadratic expansion of the eigenvalues derived in
Sec.~\ref{sec:spin-spectrum},
the total Hamiltonian can be decomposed into independent sectors labeled
by the spin eigenstates $|\varphi_{j}\rangle$. For each sector,
the Hamiltonian reduces to an effective Hamiltonian acting on the
relative vibrational coordinate:
\begin{equation}
\mathcal{H}_{j}=\frac{p^{2}}{2m}+\frac{\kappa}{2}r^{2}+\mathfrak{e}^{(0)}_{j}-r\,\mathfrak{e}^{(1)}_{j}+r^{2}\,\mathfrak{e}^{(2)}_{j}.
\end{equation}

Collecting the quadratic contributions in $r$, we introduce a spin-dependent
effective spring constant
\begin{equation}
K_{j}=\kappa+2\mathfrak{e}^{(2)}_{j}>0,\label{eq:K_eff}
\end{equation}
so that the Hamiltonian becomes
\begin{equation}
\mathcal{H}_{j}=\frac{p^{2}}{2m}+\frac{1}{2}K_{j}r^{2}-\mathfrak{e}^{(1)}_{j}r+\mathfrak{e}^{(0)}_{j}.
\end{equation}
The condition $K_{j}>0$ ensures that the effective potential remains
confining for all spin sectors. Physically, this corresponds to the
stability of the harmonic approximation, guaranteeing that dimer fluctuations
occur around a well-defined equilibrium position.

The linear and quadratic terms in $r$ can be combined by completing
the square:
\begin{equation}
\frac{1}{2}K_{j}r^{2}-\mathfrak{e}^{(1)}_{j}r=\frac{1}{2}K_{j}\left(r-\frac{\mathfrak{e}^{(1)}_{j}}{K_{j}}\right)^{2}-\frac{\bigl(\mathfrak{e}^{(1)}_{j}\bigr)^{2}}{2K_{j}}.
\end{equation}
Introducing the shifted coordinate $\tilde{r}=r-\frac{\mathfrak{e}^{(1)}_{j}}{K_{j}}$,
the Hamiltonian takes the form
\begin{equation}
\mathcal{H}_{j}=\frac{p^{2}}{2m}+\frac{1}{2}K_{j}\tilde{r}^{2}+\mathfrak{e}^{(0)}_{j}-\frac{\bigl(\mathfrak{e}^{(1)}_{j}\bigr)^{2}}{2K_{j}}.
\end{equation}

Thus, within the quadratic expansion, each spin sector is mapped onto
a shifted harmonic oscillator with renormalized stiffness $K_{j}$,
whose equilibrium position is displaced by an amount proportional
to $\mathfrak{e}^{(1)}_{j}$. The magnetoelastic coupling therefore
induces both a shift of the equilibrium coordinate and a modification
of the vibrational frequency.

Within this quadratic description, the energy spectrum of each sector
is obtained by direct quantization of the shifted harmonic oscillator,
yielding
\begin{equation}
E_{j,n}=\mathfrak{e}^{(0)}_{j}-\frac{\bigl(\mathfrak{e}^{(1)}_{j}\bigr)^{2}}{2K_{j}}+\hbar\omega_{j}\left(n+\frac{1}{2}\right).
\end{equation}
Here $\omega_j=\sqrt{K_j/m}$ is the sector-dependent vibrational
frequency.

With the sector displacement $q_j=\mathfrak e_j^{(1)}/K_j$ and thermal
width
$\sigma_j^2(T)=[\hbar/(2m\omega_j)]
\coth[\beta\hbar\omega_j/2]$, one has
$\langle r^2\rangle_{j,T}=q_j^2+\sigma_j^2(T)$, and
$\eta(T)=\max_{j,\alpha}\{|\gamma_\alpha|
\sqrt{q_j^2+\sigma_j^2(T)}\}$. Together with $K_j>0$ for every
appreciably occupied sector, this gives the stability and
small-displacement test. Explicit cubic and quartic terms additionally
require Eq.~\eqref{eq:anh-robust}. In the classical high-temperature limit,
$\sigma_j^2(T)\simeq k_BT/K_j$, giving the necessary sector-wise
fluctuation condition $k_BT\ll K_j/\gamma_\alpha^2$, which does not
constrain $q_j$.

The quantity
\begin{equation}
\varepsilon_{j}=\mathfrak{e}^{(0)}_{j}-\frac{\bigl(\mathfrak{e}^{(1)}_{j}\bigr)^{2}}{2K_{j}},\label{eq:ren-Ej}
\end{equation}
should be understood as renormalized spin energies incorporating the
static effect of dimer displacement.

Because this correction depends explicitly on $j$, the magnetoelastic
coupling modifies the energy gaps between the eigenstates $|\varphi_{j}\rangle$.
As a consequence, both the thermal populations and the occupation
of entangled states are altered. In addition, the dependence of the
vibrational frequency $\omega_{j}$ on the spin sector provides an
additional channel through which dimer fluctuations influence the
thermodynamic and quantum properties of the dimer.

\subsection{Sensitivity to cubic and quartic terms}
\label{sec:anharmonic}

To test the leading anharmonic corrections, write in sector $j$
\begin{equation}
\delta V_j(r)=a_j r^3+b_j r^4+\mathcal O(r^5),
\label{eq:anh-potential}
\end{equation}
where $a_j$ and $b_j$ include exchange and bare-elastic contributions.
This is a local perturbative expansion: a cubic truncation is not
globally stable, while a global quartic potential requires a stabilizing
quartic coefficient.

For $r=q_j+\widetilde r$, the harmonic variable $\widetilde r$ is a
zero-mean Gaussian of variance $\sigma_j^2(T)$
(Sec.~\ref{sec:reduction}), giving
\begin{equation}
\delta F_j^{(1)}
=a_j\left(q_j^3+3q_j\sigma_j^2(T)\right)
+b_j\left(q_j^4+6q_j^2\sigma_j^2(T)+3\sigma_j^4(T)\right).
\label{eq:delta-F-anh}
\end{equation}
The cubic term survives when $q_j\ne0$, whereas the quartic term grows
with the thermal width. The population correction is
\begin{equation}
\delta p_j=-\beta p_j\left(\delta F_j^{(1)}
-\sum_k p_k\delta F_k^{(1)}\right).
\label{eq:delta-p-anh}
\end{equation}
Here $p_j$ denotes the quadratic-model population; a common free-energy
shift cancels. At finite temperature the quadratic predictions are
perturbatively robust when all appreciably occupied sectors satisfy $K_j>0$,
$\eta(T)\ll1$, and
\begin{equation}
\beta\max_{j,k}\left|\delta F_j^{(1)}
-\delta F_k^{(1)}\right|\ll1.
\label{eq:anh-robust}
\end{equation}
For specified $a_j$ and $b_j$, Eqs.~\eqref{eq:delta-F-anh}--\eqref{eq:anh-robust}
directly test robustness. Corrections to concurrence, LQU, and Fisher
information follow from $p_j\to p_j+\delta p_j$, differentiating the
corrected populations with respect to $B$ for the latter.

At zero temperature, the corresponding sector-energy corrections must
remain small relative to the competing-sector gaps. At the ED--FP crossing they
shift the boundary. Outside the perturbative regime, higher-order terms
may reorder levels and shift the thermal crossover, Fisher maximum,
and LQU persistence range.

Displacement anharmonicity differs from non-Heisenberg spin
interactions. Biquadratic exchange can renormalize anisotropy, gaps,
and thermal stability in two-dimensional magnets \citep{Kartsev}, but
is absent from the present spin-$1/2$ dimer model.

\subsection{Ground-state phase diagram}
\label{sec:phase}

\begin{figure}
\centering
\includegraphics[width=0.7\linewidth]{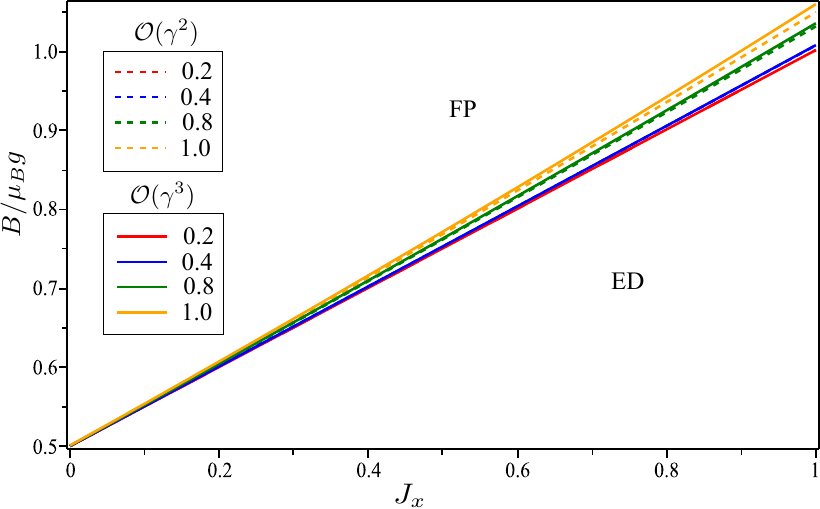}

\caption{\label{fig:0TphD}Zero-temperature phase diagram in the plane $B/\mu_{B}g$
versus $J_{x}$, for $J_{z}=-1$, $\kappa=5$, at different values
of $\gamma$. The dashed line includes only the linear magnetoelastic
correction, while the solid line incorporates both linear and quadratic
contributions.}
\end{figure}

Figure~\ref{fig:0TphD} shows the zero-temperature phase diagram
in the plane $B/\mu_{B}g$ versus $J_{x}$, for $J_{z}=-1$ and $\kappa=5$,
at different values of $\gamma=\gamma_{x}=\gamma_{z}$. Two distinct
phases are identified. At low magnetic fields, the ground state lies
in the magnetic moment $m=0$ sector and corresponds to an entangled
dimer (ED) phase, characterized by antiparallel spin correlations
and nonzero quantum entanglement. At sufficiently large fields, the
Zeeman term dominates and the system becomes fully polarized in the
product state $|^{+}_{+}\rangle$, defining the fully polarized (FP)
phase.

The phase boundary is determined by the competition between the exchange
interactions and the magnetic field. In the rigid case, the transition
corresponds to a simple level crossing. In the present magnetoelastic
system, however, the energies $\varepsilon_{j}$ are renormalized
by dimer fluctuations, which modifies the location of the critical
field. The dashed curves correspond to the phase boundary obtained
by retaining only the linear magnetoelastic correction, while the
solid curves include both linear and quadratic contributions. The
increasing deviation between these curves with $\gamma$ indicates
that quadratic magnetoelastic effects become progressively more relevant
as the exchange coupling becomes more sensitive to the dimer displacement.

Overall, Fig.~\ref{fig:0TphD} demonstrates that the magnetoelastic
coupling shifts the transition between the ED and FP phases, enlarging
or reducing the stability region of the entangled ground state depending
on the system parameters.

\subsection{Relation to coupled-dimer systems}
\label{sec:coupled-dimers}

The isolated solution is a local building block for dimer crystals.
Interdimer exchange produces dispersive, interacting triplons and may
lead to field-induced collective order \citep{GiamarchiBEC}. In
frustrated $\mathrm{SrCu_2(BO_3)_2}$, suppressed single-triplon motion
instead favors plateaus and ordered multi-triplon states
\citep{Kageyama,Miyahara}. Magnetoelasticity can modify the local gap,
interdimer couplings, and phonon-mediated forces, thereby sharpening,
splitting, or broadening the isolated-dimer crossover.

In arrays, interdimer entanglement changes the local concurrence and
LQU, while critical fluctuations may enhance susceptibility and Fisher
information. Our exact result supplies renormalized local gaps and
Boltzmann weights for strong-dimer many-body treatments, but does not
predict collective order or critical exponents. Likewise, the
dimerized or trimerized phases of decorated magnetoelastic lattices
arise from collective constraints absent here
\citep{strecka12,strecka19}.

\section{Thermal and quantum properties}
\label{sec:thermal}

In this section, we investigate the thermodynamic properties of the
magnetoelastic dimer through the effective spectrum derived in
Sec.~\ref{sec:reduction}. In particular, we analyze the partition function, the thermal
populations $p_{j}$, and the resulting magnetoelastic displacement
and fluctuations.

For Figs.~\ref{fig:r-fluct}--\ref{fig:Fisher}, we use
$J_x=k_B=\hbar=m=1$ and $\gamma_x=\gamma_z\equiv\gamma$; other
parameters are given in the captions. The spectrum is derived in
Sec.~\ref{sec:reduction}, and its anharmonic
sensitivity is discussed in Sec.~\ref{sec:anharmonic}.

\subsection{Partition function and reduced density operator}
\label{sec:partition}

After diagonalizing the spin sector and completing the square in each
channel, the Hamiltonian associated with a fixed spin eigenstate $|\varphi_{j}\rangle$
takes the form

\begin{equation}
\mathcal{H}_{j}=\frac{p^{2}}{2m}+\frac{1}{2}K_{j}\tilde{r}^{\,2}+\varepsilon_{j},\label{eq:Hj}
\end{equation}
where the effective spring constant $K_{j}$ and the renormalized
energy $\varepsilon_{j}$ are given by \eqref{eq:K_eff} and \eqref{eq:ren-Ej},
respectively. The corresponding vibrational frequency is $\omega_{j}=\sqrt{K_{j}/m}$.
Since each sector describes a shifted harmonic oscillator, its spectrum
follows directly from standard quantization, 
\begin{equation}
E_{j,n}=\varepsilon_{j}+\hbar\omega_{j}\left(n+\tfrac{1}{2}\right),\quad j=1,\dots,4,\quad n=0,1,2,\dots\label{eq:Ejn}
\end{equation}
and all thermodynamic quantities follow from these energy levels.
The total partition function becomes
\begin{equation}
\mathcal{Z}=\sum^{4}_{j=1}\sum^{\infty}_{n=0}e^{-\beta\left(\varepsilon_{j}+\hbar\omega_{j}(n+\tfrac{1}{2})\right)}=\sum^{4}_{j=1}\frac{e^{-\beta\varepsilon_{j}}}{2\sinh(\frac{\beta\hbar\omega_{j}}{2})}.\label{eq:Z}
\end{equation}

The full thermal density operator reads
\begin{equation}
\rho_{\mathrm{tot}}=\sum^{4}_{j=1}\sum^{\infty}_{n=0}P_{j,n}\,|\varphi_{j},n\rangle\langle\varphi_{j},n|,
\end{equation}
where $|\varphi_{j},n\rangle=|\varphi_{j}\rangle\otimes|n_{j}\rangle$,
with $|n_{j}\rangle$ denoting the $n$-th vibrational eigenstate
in sector $j$. The joint occupation probabilities are
\begin{equation}
P_{j,n}=\frac{1}{\mathcal{Z}}e^{-\beta(\varepsilon_{j}+\hbar\omega_{j}(n+1/2))},\label{eq:Pjn}
\end{equation}
or, equivalently, 
\begin{equation}
P_{j,n}=p_{j}\left(1-e^{-\beta\hbar\omega_{j}}\right)e^{-\beta\hbar\omega_{j}n},
\qquad \text{with} 
\qquad
p_{j}=\frac{e^{-\beta\varepsilon_{j}}}{2\mathcal{{\displaystyle Z}}\sinh(\frac{\beta\hbar\omega_{j}}{2})}.\label{eq:spop}
\end{equation}
The populations $p_{j}$ incorporate both the energy $\varepsilon_{j}$
and the vibrational contribution through $\omega_{j}$.

Tracing out the vibrational degree of freedom, one obtains the reduced
density operator of the spin dimer,

\begin{equation}
\rho=\mathrm{Tr}\rho_{\mathrm{tot}}=\frac{1}{2\mathcal{Z}}\sum^{4}_{j=1}\frac{e^{-\beta\varepsilon_{j}}}{\sinh(\frac{\beta\hbar\omega_{j}}{2})}\;|\varphi_{j}\rangle\langle\varphi_{j}|.\label{eq:rho-prj}
\end{equation}
Since the eigenstates are preserved and no off-diagonal coupling is
introduced by the magnetoelastic interaction, the reduced density
operator remains diagonal in the spin eigenstates $\{|\varphi_{j}\rangle\}$,
\begin{equation}
\rho=\sum^{4}_{j=1}p_{j}\;|\varphi_{j}\rangle\langle\varphi_{j}|.\label{eq:rrho}
\end{equation}

These expressions follow directly from the exact diagonalization of
the effective quadratic Hamiltonian in each spin sector. Since each
sector is associated with a distinct vibrational frequency $\omega_{j}$,
the partition function does not factorize into independent spin and
vibrational contributions. A factorized form is recovered when the
effective stiffness $K_{j}$ becomes independent of the spin sector,
so that all vibrational frequencies coincide.

\subsection{Magnetoelastic displacement and fluctuations}
\label{sec:displacement}

The expressions for the magnetoelastic displacement and its standard
deviation are derived in Appendix~\ref{app:displacements}.

\begin{figure}
\centering
\includegraphics[width=0.8\linewidth]{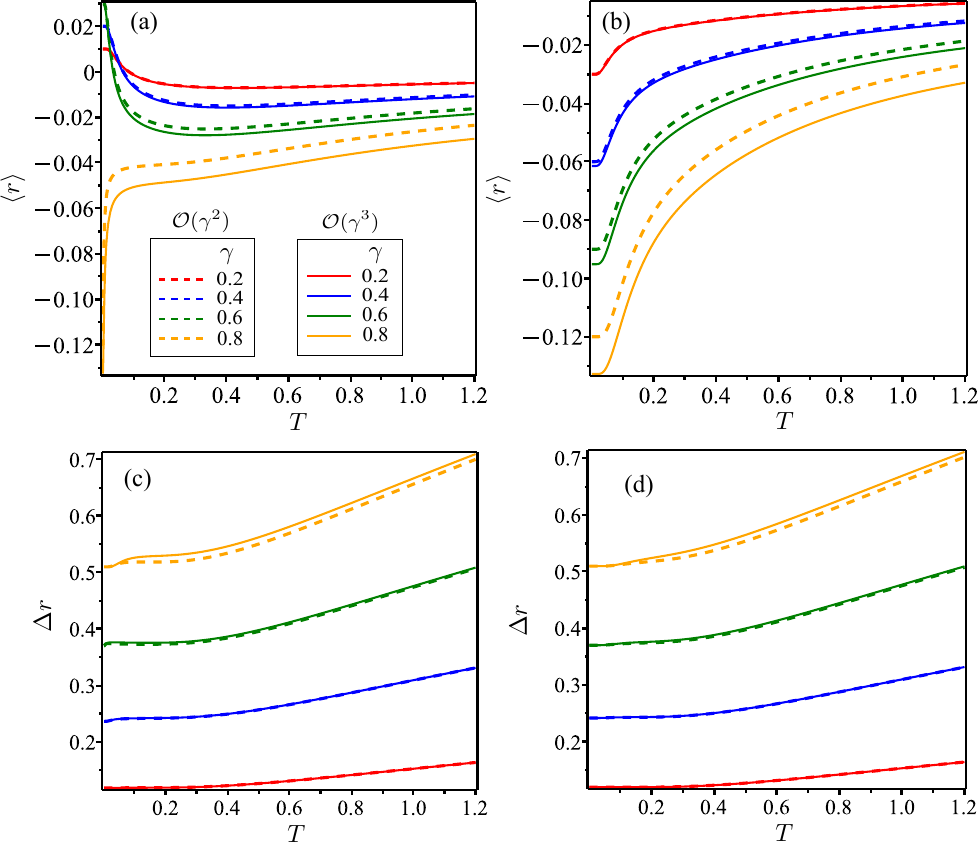}

\caption{\label{fig:r-fluct}Thermal average displacement $\langle r\rangle$
(a,b) and standard deviation $\Delta r$ (c,d) as functions of temperature
$T$, for $J_x=1$, $J_{z}=-1$ and $\kappa=5$, at different values of $\gamma$.
Dashed curves include only the linear magnetoelastic contribution,
while solid curves include both linear and quadratic corrections.
Panels (a,c) correspond to $B/\mu_{B}g=1.1$, and (b,d) to $B/\mu_{B}g=0.9$.}
\end{figure}

Figure~\ref{fig:r-fluct} shows the thermal average displacement
$\langle r\rangle$ given in \eqref{eq:agv-r} and the standard deviation
$\Delta r$ [Eq.~\eqref{eq:std-dev}] as functions of temperature
$T$, for two representative magnetic fields located on opposite sides
of the zero-temperature phase boundary. Panels (a,c), with $B/\mu_{B}g=1.1$,
correspond to the fully polarized (FP) region, while panels (b,d),
with $B/\mu_{B}g=0.9$, probe the entangled dimer (ED) regime.

At low temperatures, $\langle r\rangle$ approaches a constant value
determined by the ground state. As the temperature increases, thermal
population of excited levels produces a smooth crossover between distinct
spin sectors, an effect that is more pronounced in the ED regime due
to the smaller energy separation between competing states.

The behavior of $\Delta r$ further highlights this competition. In
both regimes, $\Delta r$ increases with temperature as vibrational
excitations become thermally activated, although the enhancement is
significantly stronger in the ED phase, indicating larger fluctuations
associated with nearly degenerate spin configurations. The comparison
between dashed and solid curves also shows that the quadratic magnetoelastic
correction modifies both $\langle r\rangle$ and $\Delta r$, particularly
the fluctuation amplitude.

Besides characterizing the lattice fluctuations, this analysis
provides a partial fluctuation-based check through
$|\gamma_{\alpha}|\Delta r\ll1$. The full sector-resolved
small-displacement test is $\eta(T)\ll1$ [Eq.~\eqref{eq:eta-validity}],
which also includes the sector displacements $q_j$. The results also
reveal how thermal vibrational effects influence the renormalized
spectrum and the resulting quantum correlations discussed below.

\subsection{Level populations and thermal redistribution}
\label{sec:populations}

\begin{figure}
\centering
\includegraphics[width=0.8\linewidth]{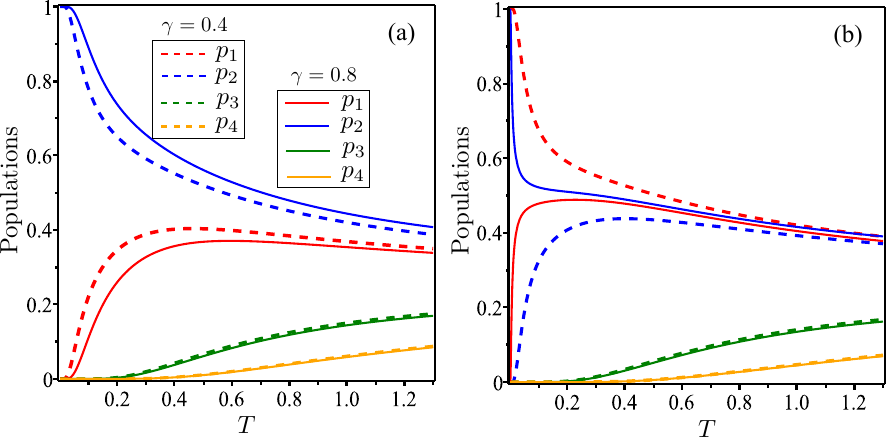}

\caption{\label{fig:popl-lev}Level populations $p_{j}$ as functions of temperature
$T$, for $J_x=1$, $J_{z}=-1$ and $\kappa=5$. Dashed curves correspond to
$\gamma=0.4$, and solid curves to $\gamma=0.8$. Panels (a) and (b)
show $B/\mu_{B}g=1.1$ and $B/\mu_{B}g=0.9$, respectively.}
\end{figure}

Figure~\ref{fig:popl-lev} shows the temperature dependence of the
level populations $p_{j}$ for two representative magnetic fields,
$B/\mu_{B}g=1.1$ (panel a) and $B/\mu_{B}g=0.9$ (panel b), and for
two values of the magnetoelastic parameter $\gamma$. These fields
probe, respectively, the fully polarized (FP) and entangled dimer
(ED) regimes identified in Fig.~\ref{fig:0TphD}.

At low temperatures, the populations are dominated by the ground state
of each regime. In the FP phase, a single level is nearly fully occupied,
while in the ED regime the low-energy states remain closer in energy,
producing a more balanced population distribution. As the temperature
increases, thermal excitations progressively populate higher-energy
levels, leading to a smooth redistribution of $p_{j}$.

The effect of the magnetoelastic coupling is evident from the comparison
between different values of $\gamma$. Increasing $\gamma$ modifies
the relative spacing between the renormalized energies $\varepsilon_{j}$,
changing the temperature scale at which excited states become significantly
populated. This effect is particularly pronounced in the ED regime,
where the competing levels are nearly degenerate.

\section{Quantum correlations and magnetic response}
\label{sec:quantum-response}

In this section, we analyze how the magnetoelastic coupling affects
quantum correlations and magnetic response through the population
rearrangement among spin sectors $p_{j}$. For this purpose, we investigate
the concurrence, the local quantum uncertainty, and the magnetic Fisher
information.

\subsection{Concurrence}
\label{sec:concurrence}

To quantify the bipartite entanglement of the thermal state, we employ
the concurrence\citep{wootters,hill}, which for a two-qubit density
operator $\rho$ is defined as
\begin{equation}
\mathcal{C}(\rho)=\max\left\{ 0,\,\lambda_{1}-\lambda_{2}-\lambda_{3}-\lambda_{4}\right\} ,
\end{equation}
 where $\lambda_{i}$ are the square roots, in decreasing order, of
the eigenvalues of the matrix
\begin{equation}
\mathcal{R}=\rho\,(\sigma_{y}\otimes\sigma_{y})\,\rho^{*}\,(\sigma_{y}\otimes\sigma_{y}).
\end{equation}

In the present case, the reduced density operator obtained in
Eq.~\eqref{eq:rho-prj}
has the block diagonal form when written in the natural basis $\{\,|^{+}_{+}\rangle,\;|^{+}_{-}\rangle,\;|^{-}_{+}\rangle,\;|^{-}_{-}\rangle\,\}$.

Using the eigenstates introduced in Eq.~\eqref{eq:phi-st}, it can
be written explicitly as 
\begin{equation}
\rho=\begin{pmatrix}p_{1} & 0 & 0 & 0\\[2mm]
0 & \dfrac{p_{2}+p_{3}}{2} & \dfrac{p_{2}-p_{3}}{2} & 0\\[2mm]
0 & \dfrac{p_{2}-p_{3}}{2} & \dfrac{p_{2}+p_{3}}{2} & 0\\[2mm]
0 & 0 & 0 & p_{4}
\end{pmatrix},\label{eq:rho-Mtx}
\end{equation}
where the populations $p_{j}$ are given by Eq.~\eqref{eq:spop}.

For a block diagonal form, the concurrence reduces to $\mathcal{C}=2\max\left\{ 0,\,|\rho_{23}|-\sqrt{\rho_{11}\rho_{44}}\right\} $,
which, in terms of populations, yields
\begin{equation}
\mathcal{C}=\max\left\{ 0,\,|p_{2}-p_{3}|-2\sqrt{p_{1}p_{4}}\right\}.
\end{equation}

This expression shows that entanglement is governed by the competition
between the population imbalance of the entangled states $|\varphi_{2}\rangle$
and $|\varphi_{3}\rangle$, and the thermal occupation of the separable
sectors $|\varphi_{1}\rangle$ and $|\varphi_{4}\rangle$. Therefore,
magnetoelastic effects influence the concurrence indirectly through
the redistribution of the thermal populations $p_{j}$.

\begin{figure}
\centering
\includegraphics[width=0.8\linewidth]{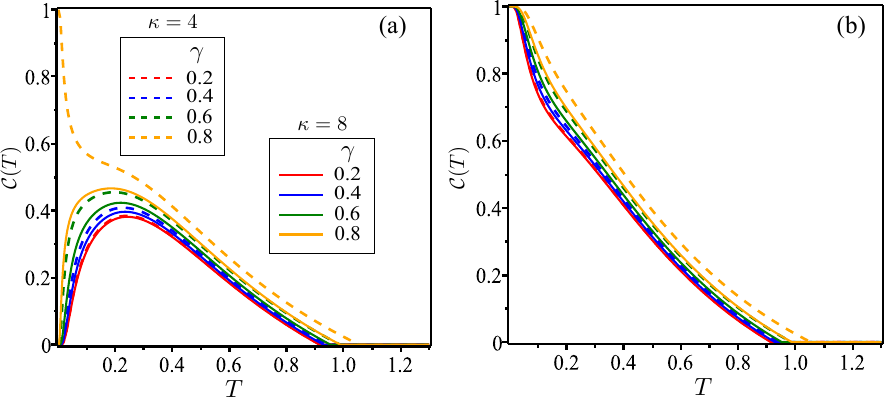}

\caption{\label{fig:conrr}Concurrence $\mathcal{C}(T)$ as a function of temperature
$T$, for $J_x=1$, $J_{z}=-1$ and different values of $\gamma$. Dashed curves
correspond to $\kappa=4$, and solid curves to $\kappa=8$. Panels
(a) and (b) show $B/\mu_{B}g=1.1$ and $B/\mu_{B}g=0.9$, respectively.}
\end{figure}

Figure~\ref{fig:conrr} shows the concurrence $\mathcal{C}$
as a function of temperature $T$, for two values of the stiffness
$\kappa$ and different values of the magnetoelastic parameter $\gamma$,
at fixed magnetic fields corresponding to the FP and ED regions identified
in Fig.~\ref{fig:0TphD}. Panels (a) and (b) probe $B/\mu_{B}g=1.1$
(FP regime) and $B/\mu_{B}g=0.9$ (ED regime), respectively. At low
temperatures, the concurrence reflects the nature of the ground state
in each regime. In the FP phase, the ground state is essentially separable
and $\mathcal{C}\approx0$, whereas in the ED regime the entangled
ground state produces a finite concurrence as $T\to0$. Increasing
temperature leads to thermal redistribution of the populations $p_{j}$,
causing a gradual suppression of $\mathcal{C}$.

The magnetoelastic coupling strongly affects this behavior through
the renormalized spectrum $\varepsilon_{j}$. Increasing $\gamma$
shifts the temperature scale at which concurrence is suppressed, while
the stiffness $\kappa$ controls the magnitude of the vibrational
fluctuations and consequently the strength of the magnetoelastic renormalization.
These effects are particularly pronounced in the ED regime, where
competing levels remain close in energy.

\subsection{Local quantum uncertainty}
\label{sec:lqu}

We now characterize quantum correlations beyond entanglement by means
of the local quantum uncertainty (LQU)\citep{Girolami}. This quantity
captures the minimum quantum uncertainty associated with local measurements
and remains nonzero even in separable states.

Due to the symmetry of the Hamiltonian under exchange of the two spins,
the reduced density matrix is invariant under permutation of the subsystems.
As a consequence, the LQU is identical for measurements on either
spin, i.e., $\mathcal{U}_{1}=\mathcal{U}_{2}$.

For a bipartite system, the LQU with respect to measurements on the
first spin is defined as 
\begin{equation}
\mathcal{U}_{1}(\rho)=1-\lambda_{\max}(W),
\end{equation}
where $\lambda_{\max}(W)$ is the largest eigenvalue of the $3\times3$
matrix
\begin{equation}
W_{ij}={\rm Tr}\left[\sqrt{\rho}(\sigma_{i}\otimes\mathbb{I})\sqrt{\rho}(\sigma_{j}\otimes\mathbb{I})\right],\quad i,j=x,y,z.
\end{equation}
In the present model, the reduced density operator is diagonal in
the eigenstates $\{|\varphi_{j}\rangle\}$, as given in
Eq.~\eqref{eq:rrho},
$\rho=\sum^{4}_{j=1}p_{j}|\varphi_{j}\rangle\langle\varphi_{j}|$,
so that $\sqrt{\rho}=\sum^{4}_{j=1}\sqrt{p_{j}}\,|\varphi_{j}\rangle\langle\varphi_{j}|$.

Using this structure, the matrix $W$ can be evaluated analytically.
For the block diagonal form of the density matrix in
Eq.~\eqref{eq:rho-Mtx},
one finds that $W$ is diagonal,
\begin{equation}
W=\begin{pmatrix}W_{x} & 0 & 0\\
0 & W_{x} & 0\\
0 & 0 & W_{z}
\end{pmatrix},
\end{equation}
with the explicit coefficients 
\begin{alignat}{1}
W_{x}=  (\sqrt{p_{1}}+\sqrt{p_{4}})(\sqrt{p_{2}}+\sqrt{p_{3}}),\qquad
W_{z}=  p_{1}+p_{4}+2\sqrt{p_{2}p_{3}}.
\end{alignat}

The quantity $W_{x}$ couples populations belonging to different magnetic
sectors, while $W_{z}$ is particularly sensitive to the coexistence
of the two $m=0$ states through the geometric contribution $\sqrt{p_{2}p_{3}}$.

Therefore, the LQU reduces to the simple form 
\begin{equation}
\mathcal{U}_{1}=1-\max\left(W_{x},W_{z}\right).\label{eq:Ua}
\end{equation}
Analytical expressions for LQU and related Fisher-based quantifiers
in X states have also been discussed in Refs. \citep{Fedorova,Yurischev}.

The LQU depends exclusively on the thermal populations $p_{j}$, which
are determined by the magnetoelastic spectrum through
Eq.~\eqref{eq:spop}.
In particular, it reflects the competition between the entangled states
$|\varphi_{2}\rangle$, $|\varphi_{3}\rangle$ and the separable states
$|\varphi_{1}\rangle$, $|\varphi_{4}\rangle$.

Unlike concurrence, the LQU may remain finite even when thermal mixing
suppresses entanglement,
\begin{equation}
\mathcal{C}=0,\quad\text{while}\quad\mathcal{U}_{1}>0,
\end{equation}
revealing the persistence of nonclassical correlations. In the present
magnetoelastic system, this behavior typically occurs when $p_{2}$
and $p_{3}$ become nearly equal, suppressing concurrence while preserving
finite quantum correlations induced by the interplay between magnetic
and vibrational degrees of freedom.

\begin{figure}
\centering
\includegraphics[width=0.8\linewidth]{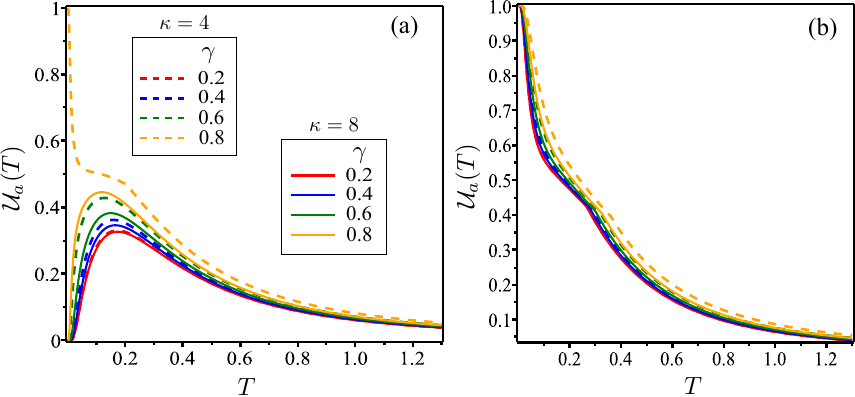}

\caption{\label{fig:ldu}Local quantum uncertainty $\mathcal{U}_{1}(T)$ as
a function of temperature $T$, for $J_x=1$, $J_{z}=-1$ and different values
of $\gamma$. Dashed curves correspond to $\kappa=4$, and solid curves
to $\kappa=8$. Panels (a) and (b) show $B/\mu_{B}g=1.1$ and $B/\mu_{B}g=0.9$,
respectively.}
\end{figure}

Figure~\ref{fig:ldu} shows the temperature dependence of
the local quantum uncertainty $\mathcal{U}_{1}(T)$ for two representative
magnetic fields located on opposite sides of the zero-temperature
phase boundary, and for different values of the magnetoelastic parameters
$\gamma$ and $\kappa$.

At low temperatures, the behavior of $\mathcal{U}_{1}$ reflects the
nature of the ground state in each regime. In the fully polarized
(FP) phase, shown in panel (a), the ground state is essentially separable,
with $p_{1}\approx1$, leading to $W_{x}\approx0$ and $W_{z}\approx1$,
and consequently $\mathcal{U}_{1}\approx0$. In contrast, in the entangled
dimer (ED) regime shown in panel (b), the populations $p_{2}$ and
$p_{3}$ become relevant, producing finite contributions through the
term $\sqrt{p_{2}p_{3}}$ and resulting in a nonzero LQU at low temperatures.

As the temperature increases, thermal redistribution of the populations
$p_{j}$ produces a nonmonotonic behavior of $\mathcal{U}_{1}(T)$,
with a maximum at intermediate temperatures where several spin sectors
contribute simultaneously. The magnetoelastic coupling strongly affects
this behavior through the renormalized spectrum $\varepsilon_{j}$
and the vibrational frequencies $\omega_{j}$, shifting both the position
and amplitude of the maximum. In particular, larger values of $\gamma$
broaden the temperature region where competing spin sectors coexist.

\subsection{Magnetic Fisher information}
\label{sec:fisher}

To characterize the sensitivity of the system to variations of the
external magnetic field, we introduce the Fisher information\citep{Braunstein,paris}
associated with the parameter $B$. This quantity measures how strongly
the thermal state responds to infinitesimal changes of the magnetic
field and therefore quantifies the field sensitivity of the system.

Since the eigenstates of the reduced density operator are independent
of $B$, the quantum Fisher information reduces exactly to the classical
Fisher information associated with the thermal populations $p_{j}(B)$.
Consequently, the Fisher information is given by
\begin{equation}
\mathcal{F}_{B}=\sum^{4}_{j=1}\frac{1}{p_{j}}\left(\frac{\partial p_{j}}{\partial B}\right)^{2}.\label{eq:FB}
\end{equation}

In the present model, the populations are given by \eqref{eq:spop},
\begin{equation}
p_{j}=\frac{e^{-\beta\varepsilon_{j}(B)}}{2\mathcal{Z}\sinh(\beta\hbar\omega_{j}/2)},
\end{equation}
where the dependence on the magnetic field enters through the spin
energies $\varepsilon_{j}(B)$, while the vibrational frequencies
$\omega_{j}$ do not depend on $B$.

It is convenient to rewrite Eq.~\eqref{eq:FB} in a more compact and
physically transparent form. Taking the derivative of $p_{j}$ and
using normalization, one obtains
\begin{equation}
\frac{\partial p_{j}}{\partial B}=-\beta\,p_{j}\left[\frac{\partial\varepsilon_{j}}{\partial B}-\left\langle \frac{\partial\varepsilon}{\partial B}\right\rangle \right],\label{eq:pB}
\qquad\text{where} \quad
\left\langle \frac{\partial\varepsilon}{\partial B}\right\rangle =\sum^{4}_{k=1}p_{k}\frac{\partial\varepsilon_{k}}{\partial B}.
\end{equation}
Substituting Eq.~\eqref{eq:pB} into Eq.~\eqref{eq:FB}, the Fisher
information reduces to
\begin{equation}
\mathcal{F}_{B}=\beta^{2}\left[\langle m^{2}\rangle-\langle m\rangle^{2}\right],\label{eq:FBm}
\end{equation}
where $m_{j}=-\partial\varepsilon_{j}/\partial B$ is the magnetic
moment associated with the eigenstate $|\varphi_{j}\rangle$. Therefore,
$\mathcal{F}_{B}$ measures the thermal fluctuations of the magnetization
and provides a direct probe of the magnetic response of the system.
Through the fluctuation-response relation, this quantity is proportional
to the magnetic susceptibility.

Generally, QFI contains eigenvalue and eigenvector contributions. Here
the field commutes with the Hamiltonian and all eigenvectors are
$B$-independent. Thus Eq.~\eqref{eq:FB} is the full equilibrium QFI for
both the reduced spin state and, because the conditional oscillator
states are also $B$-independent, the full spin--vibrational Gibbs state.
For $\nu$ independent repetitions,
\begin{equation}
\delta B\geq\frac{1}{\sqrt{\nu\mathcal F_B}},
\end{equation}
while Eq.~\eqref{eq:FBm} gives $\mathcal F_B=\beta\chi_B$ for
$\chi_B=\partial\langle m\rangle/\partial B$.

Multipartite-entanglement witnesses instead use QFI generated by a
collective spin and compare it with $k$-producibility bounds
\citep{Hauke,MathewQFI,Pezzereview}; it may become singular or
superextensive at a many-body critical point. The finite dimer has only
a rounded thermal peak at its ED--FP level crossing. Because the
present QFI is population-based, a large value is not itself an
entanglement witness. Magnetoelasticity tunes the peak, but independent
dimers have additive QFI; beyond-standard-quantum-limit scaling would
require collective correlations and an explicit resource benchmark.

\begin{figure}[H]
\centering
\includegraphics[width=0.8\linewidth]{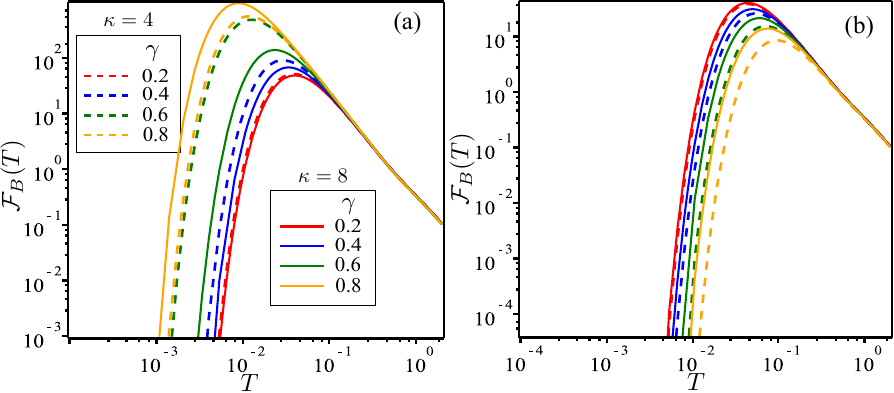}

\caption{\label{fig:Fisher}Magnetic Fisher information $\mathcal{F}_{B}(T)$
as a function of temperature $T$, for $J_x=1$, $J_{z}=-1$ and different values
of $\gamma$. Dashed curves correspond to $\kappa=4$, and solid curves
to $\kappa=8$. Panels (a) and (b) show $B/\mu_{B}g=1.1$ and $B/\mu_{B}g=0.9$,
respectively.}
\end{figure}

Figure~\ref{fig:Fisher} illustrates the temperature dependence of
the magnetic Fisher information $\mathcal{F}_{B}(T)$ shown on a logarithmic
scale for the same set of parameters considered in the previous figures,
allowing a direct comparison with the behavior of concurrence and
LQU.

At low temperatures, the behavior of $\mathcal{F}_{B}$ reflects the
nature of the ground state. In the fully polarized (FP) regime, the
magnetization is essentially fixed and magnetic fluctuations vanish,
leading to $\mathcal{F}_{B}\to0$ as $T\to0$. In contrast, in the
entangled dimer (ED) regime, low-lying states with different magnetic
moments contribute to finite fluctuations, producing a rapid increase
of $\mathcal{F}_{B}$ at small temperatures.

A prominent feature of Fig.~\ref{fig:Fisher} is the appearance of
pronounced maxima at intermediate temperatures, signaling regions
where the populations $p_{j}$ become highly sensitive to the magnetic
field. Microscopically, this occurs when competing levels with distinct
magnetic slopes $\partial\varepsilon_{j}/\partial B$ coexist, leading
to strong redistribution of populations under small variations of
$B$. The position and amplitude of these peaks are strongly affected
by the magnetoelastic parameters $\gamma$ and $\kappa$, which modify
both the renormalized spectrum and the effective magnetic moments.

Unlike concurrence and LQU, the Fisher information measures
field sensitivity rather than intrinsic quantum correlations; it can
therefore remain enhanced after those correlations are thermally
suppressed.

\subsection{Experimental observables and protocol}
\label{sec:experiment}

The population mechanism can be tested in strongly dimerized materials
by combining spectroscopy and thermodynamic response. Neutron
scattering or ESR measures the field-dependent singlet--triplet gap,
while Raman scattering tracks dimer-sensitive phonons and the predicted
sector-dependent $\omega_j$. In $\mathrm{SrCu_2(BO_3)_2}$, pantograph
modes modulate intradimer exchange \citep{Thirunavukkuarasu}, and their
pressure-dependent shift follows dimer spin correlations
\citep{Bettler}. On a fine $(B,T)$ grid, static susceptibility, or
sufficiently low-frequency ac susceptibility in the equilibrium limit,
gives $\mathcal F_B=\beta\chi_B$. Magnetostriction, sound velocity, and
thermal expansion additionally probe $\langle r\rangle$, elastic
softening, and Eq.~\eqref{eq:eta-validity}.

Spectral energies and intensities constrain $p_j$ through
Eq.~\eqref{eq:spop}; neutron spin correlators, supplemented by NMR,
test their redistribution. With the symmetry of
Eq.~\eqref{eq:rho-Mtx}, correlators and magnetization reconstruct the
two-spin state and hence concurrence and LQU. This comparison is most
quantitative for dimer compounds: although
$\mathrm{Tb_{0.3}Dy_{0.7}Fe_{1.9}}$ has strong spin--lattice response,
its itinerant, many-spin, and domain physics lie beyond this model.

\section{Conclusions}
\label{sec:conclusions}

In this work, we have presented a comprehensive analysis of a magnetoelastic
spin-1/2 Heisenberg dimer, highlighting the role of dimer fluctuations
in shaping both thermodynamic and quantum properties. The exact treatment
of the spin sector, combined with a harmonic description of the vibrational
degree of freedom, leads to an effective model in which each spin
configuration is associated with a distinct vibrational mode. This
structure results in a non-factorizable partition function and induces
a nontrivial redistribution of thermal populations.

Our results show that the physical behavior of the system is governed
by the competition between spin sectors, which is directly controlled
by the magnetoelastic renormalization of the energy spectrum. This
mechanism provides a unified explanation for the behavior of different
physical quantities. In particular, the concurrence captures the presence
of entanglement and sharply distinguishes between entangled and fully
polarized regimes, while the LQU reveals more robust nonclassical
correlations that persist beyond the entanglement threshold.

From a complementary perspective, the magnetic Fisher information
probes the response of the system to external perturbations and highlights
regions of enhanced sensitivity. Its behavior reflects the same underlying
competition between energy levels, establishing a direct connection
between quantum statistical fluctuations and magnetic response. In
this sense, entanglement, LQU, and Fisher information provide complementary
signatures of the same magnetoelastic redistribution mechanism.

Overall, the present results demonstrate that magnetoelastic coupling
offers a natural route to control both quantum correlations and thermodynamic
response in low-dimensional systems. The unified description provided
here suggests that the interplay between spin and dimer degrees of
freedom can be systematically exploited to tune nonclassical properties
and enhance sensitivity to external fields. These findings open perspectives
for extending the analysis to more complex spin-dimer systems, where
collective effects may further amplify the role of magnetoelastic
interactions.

These conclusions require $K_j>0$ in appreciably occupied sectors,
$\eta(T)\ll1$ [Eq.~\eqref{eq:eta-validity}], and the anharmonic
condition \eqref{eq:anh-robust} (or corrections smaller than sector
gaps at zero temperature). Outside this window, higher-order terms
require nonperturbative treatment. The single-dimer Fisher peak is a
finite population crossover, not evidence by itself of multipartite
entanglement or beyond-standard-quantum-limit scaling. Coupled dimers
may instead exhibit triplon dispersion, collective phases, and critical
enhancement.

\section*{Acknowledgements}

This work was partially supported by Brazilian agencies CNPq and Fapemig. M.R.
acknowledges CNPq Grant No.~311565/2025-5.

\begin{appendices}
\renewcommand{\theHequation}{A.\arabic{equation}}

\section{Displacements and standard deviation}
\label{app:displacements}

To assess the temperature regime where the harmonic approximation
remains valid, we analyze the thermal averages of the linear and quadratic
displacements. The associated standard deviation provides a direct
measure of dimer fluctuations.

\subsection{Linear displacement}
\label{app:linear-displacement}

From Eq.~\eqref{eq:phi-st}, for each spin eigenstate $|\varphi_{j}\rangle$
the shifted coordinate is 
\begin{equation}
\tilde{r}=r-\frac{\mathfrak{e}^{(1)}_{j}}{K_{j}},\qquad K_{j}=\kappa+2\mathfrak{e}^{(2)}_{j},
\end{equation}
with frequency $\omega_{j}=\sqrt{K_{j}/m}$. In thermal equilibrium,
the shifted harmonic oscillator satisfies $\langle\tilde{r}\rangle_{j}=0$.
Therefore, for a fixed spin sector,
\begin{equation}
\langle r\rangle_{j}=\frac{\mathfrak{e}^{(1)}_{j}}{K_{j}}.
\end{equation}
Averaging over all spin sectors with weights $p_{j}$, one obtains
\begin{equation}
\langle r\rangle=\sum^{4}_{j=1}p_{j}\,\langle r\rangle_{j}=\sum^{4}_{j=1}p_{j}\,\frac{\mathfrak{e}^{(1)}_{j}}{K_{j}},\label{eq:agv-r}
\end{equation}
where $p_{j}$ is given by \eqref{eq:rrho}.

\subsection{Standard deviation}
\label{app:standard-deviation}

Using the relation $r=\tilde{r}+\mathfrak{e}^{(1)}_{j}/K_{j}$, the
mean square displacement in sector $j$ is
\begin{equation}
\langle r^{2}\rangle_{j}=\langle\tilde{r}{}^{2}\rangle_{j}+\left(\frac{\mathfrak{e}^{(1)}_{j}}{K_{j}}\right)^{2},
\end{equation}
since $\langle\tilde{r}\rangle_{j}=0$. For a harmonic oscillator
of frequency $\omega_{j}$,
\begin{equation}
\langle\tilde{r}{}^{2}\rangle_{j}=\frac{\hbar}{2m\omega_{j}}\coth\left(\frac{\beta\hbar\omega_{j}}{2}\right).
\end{equation}

Therefore, the total mean square displacement is 
\begin{equation}
\langle r^{2}\rangle=\sum^{4}_{j=1}p_{j}\left[\frac{\hbar}{2m\omega_{j}}\coth\left(\frac{\beta\hbar\omega_{j}}{2}\right)+\left(\tfrac{\mathfrak{e}^{(1)}_{j}}{K_{j}}\right)^{2}\right].
\end{equation}

The standard deviation of the displacement is 
\begin{equation}
\Delta r=\sqrt{\langle r^{2}\rangle-\langle r\rangle^{2}}.\label{eq:std-dev}
\end{equation}

\subsection{Physical interpretation}
\label{app:physical-interpretation}

The displacement $\langle r\rangle$ represents a thermally weighted
shift of the equilibrium position induced by the magnetoelastic coupling.
The fluctuations $\Delta r$ quantify the spread of the dimer coordinate
around this shifted equilibrium.

The full small-displacement test is $\eta(T)\ll1$
[Eq.~\eqref{eq:eta-validity}], which includes both $q_j$ and
$\sigma_j(T)$ in every relevant sector. Thus
$|\gamma_{\alpha}|\Delta r\ll1$ is only a partial fluctuation check.
Explicit cubic and quartic terms also require
Eq.~\eqref{eq:anh-robust}; if either condition fails, the quadratic
predictions need not remain reliable.

At low temperatures, 
\[
\langle\tilde{r}^{2}\rangle_{j}\to\frac{\hbar}{2m\omega_{j}},
\]
so the fluctuations are dominated by zero-point motion. At high temperatures,
\[
\langle\tilde{r}^{2}\rangle_{j}\sim\frac{k_{B}T}{K_{j}},
\]
leading to classical thermal fluctuations that grow linearly with
temperature. The sector dependence of $K_{j}$ implies that different
spin configurations contribute unequally to the total fluctuations,
providing a direct link between magnetic and dimer degrees of freedom.
\end{appendices}

\end{document}